# Enhanced gyration-signal propagation speed in one-dimensional vortex-antivortex lattices and its control by perpendicular bias field

Sang-Koog Kim[†,*] and Han-Byeol Jeong[†]

*National Creative Research Initiative Center for Spin Dynamics and Spin-Wave Devices, Nanospinics Laboratory, Research Institute of Advanced Materials, and Department of Materials Science and Engineering, Seoul National University, Seoul 151-744, Republic of Korea*

We report on a micromagnetic simulation study of coupled core gyrations in one-dimensional (1D) alternating vortex-antivortex (V-AV) lattices formed in connected soft-magnetic-disk arrays. In such V-AV lattices, we found very characteristic standing-wave gyration modes as well as ultrafast gyration-signal propagation, as originating from combined strong exchange and dipole interactions between the neighboring vortices and antivortices. Collective core oscillations in the V-AV networks are characterized as unique two-branch bands that are affected by the vortex-antivortex polarization ordering and controllable by externally applied perpendicular fields. The gyration-signal propagation speed is much faster than that for 1D disk arrays composed only of vortex states, and the propagation speed for the parallel polarization ordering is increased, remarkably, to more than 1 km/s by application of perpendicular static fields. This work provides a fundamental understanding of the coupled dynamics of topological solitons as well as an additional mechanism for ultrafast gyration-signal propagation; moreover, it offers an efficient means of significant propagation-speed enhancement that is suitable for information carrier applications in continuous thin-film nanostrips.

Nontrivial spin textures, especially the magnetic vortex in confined potential wells such as ferromagnetic nanodots, have been widely studied owing not only to their intriguing gyration dynamics[1-6] and dynamic core-switching behaviors[7-12] but also their potential applications for information-storage[13-16] and -processing[17-20] as well as microwave devices.[21,22] Very recently, studies on the dynamics of single vortices have been extended to their coupled systems, such as two-vortex pairs[18, 19] and one- or two-dimensional (1D or 2D) periodic arrays of coupled vortices,[17,20,23,24] in order to explore the fundamental modes of coupled vortex gyrations and their controllability.

In the meantime, the magnetic antivortex, the topological counterpart of the vortex, has been found to exhibit core gyration and related switching behaviors similar to vortex dynamics.[25-27] Periodic vortex and antivortex arrangements have often been found as parts of cross-tie walls.[28,29] Isolated antivortices by contrast, due to their unstable state, possibly can be formed in specially designed geometric confinements.[25-27,29] The antivortex also appears together with vortices in dynamic transient states during vortex-core reversals in nanodots[7-11] and magnetization-reversal dynamics,[30] domain-wall motions in nanostrips[31,32] or highly nonlinear chaos dynamics,[33] because the total topological charge is always conserved during such dynamic phenomena.[34,35] Although the dynamics of single antivortices[25-27] and their dynamic interactions with vortices[30,36,37] have been reported in earlier studies, the dynamics of coupled vortices and antivortices in round-shaped modulated nanostrips and their coupled gyration propagation have not been well understood in the magnonic band aspect or in terms of gyration-signal propagations through 1D alternating vortex-antivortex (V-AV) lattices.

In this letter, we report on the fundamental dynamic behaviors of interacting vortices and antivortices in 1D V-AV lattices, specifically in terms of their coupled excitation modes, very unique two-branch band structures, and gyration-signal propagation speed. We found that gyration-signal propagations are much faster (i.e., more than 1 km/s) than those for 1D disk arrays composed only of vortex states, as originating from the strong exchange-dipole interactions between neighboring vortices and antivortices. Also, we discovered that the unique band structure and gyration-propagation speed are controllable according to the strength and direction of perpendicular static fields. This study provides not only fundamental insights into the dynamic interactions between different types of topological solitons but represents the first step toward their implementation in spin-based information signal-processing devices.

In the present study, we used, for a model system, alternating V-AV lattices in a connected triple-disk structure wherein the diameter of each disk is $2R = 303$ nm, the thickness $T = 20$ nm, and the center-to-center interdistance $D_{int} = 243$ nm, for cases of $R < D_{int} < 2R$ (see Fig. 1). Unlike physically separated disks (i.e., $D_{int} > 2R$), this connected triple-disk structure has two antivortices between neighboring vortices of the same counter-clockwise (CCW) chirality in a metastable state. In the initial state, the core magnetizations of the three vortices are upward (called polarization $p = +1$), and those of the two antivortices are downward ($p = -1$), resulting in the antiparallel polarization ordering. We numerically calculated the motions of local magnetizations (cell size: $3\times3\times T$ nm$^3$) using the OOMMF code,[38] which employs the Landau-Liftshitz-Gilbert equation.[39] We used the typical material parameters of Permalloy (Ni$_{80}$Fe$_{20}$, Py): saturation magnetization $M_s=8.6\times10^5$ A/m, exchange stiffness $A_{ex}= 1.30\times10^{-11}$ J/m, and zero magnetocrystalline anisotropy.

For excitation of all of the coupled modes in such a model system, we applied a static local field of 200 Oe in the + y direction only in the left-end disk, so as to displace the core position to ~54 nm in the −x direction. Upon turning off the local field, the five individual cores of the vortices and antivortices were monitored to extract the trajectories of the coupled core motions under free relaxation. All of the simulation results noted hereafter were obtained up to 200 ns after the field was turned off.

Figure 2 (a) shows the individual trajectories of coupled core gyrations with their position vectors **X**= (*X, Y*). Owing to direct excitation of the core in the first disk (noted as $V_1$), a large amplitude of the gyration of the core, starting at *X*= - 54 nm toward the center position, is observed. The vortex gyration of the first disk is then propagated to the next antivortex ($AV_2$), and then further propagates through the whole system, as evidenced by the large gyration amplitudes of the remaining vortices, indicated as $V_3$ and $V_5$. These large gyration amplitudes imply that the vortex gyration is well propagated to the next vortices through the neighboring antivortices, with negligible energy loss. To elucidate the observed modes, we plotted frequency spectra in the left column of Fig. 2(b), as obtained from fast Fourier transformations (FFTs) of the *x* components of the individual core oscillations. From the first vortex to the last one, contrasting FFT powers between the major five peaks are observed. The distinct peaks are denoted $\omega_i$, with *i* = 1, 2, 3, 4, and 5. The individual peaks in the frequency domains correspond to $\omega/2\pi$ = 0.23, 0.45, 0.55, 0.98, and 1.11 GHz, respectively, spreading in a wide range from a few hundreds of MHz to the GHz level. Each peak in each frequency spectrum is located at the same corresponding frequency from $V_1$ through $V_5$, indicating that there are five distinct modes in the whole system. Because of the intrinsic damping of each core gyration, those peaks are somewhat broadened and overlapped with the neighboring peaks. There are noticeable differences between the vortex and antivortex-core motions: in antivortex motions, that is, in the $AV_2$ and $AV_4$ spectra, the $\omega_3$ peak disappears, while in the V$_3$ spectrum, the $\omega_2$ peak disappears and the $\omega_4$ peak becomes very weak. In order to obtain those data with a better spectral resolution, we carried out further simulations with the same model but with an extremely low damping constant, α = 0.0001. The resultant FFT spectra (right column of Fig. 2(b)) show that the five major peaks become sharper. Moreover, for cases of antivortex gyration motions, the higher-frequency $\omega_4$ and $\omega_5$ modes, which were very weak for α = 0.01, are much stronger for α = 0.0001, due not only to the negligible energy loss but also to the strong exchange interaction between the antivortices and the neighboring vortices. The very small subsidiary peaks in the FFT spectra might be associated with higher harmonics and the nonlinear effect of the complex geometric confinements (e.g., the notch-shaped boundaries) or the discreteness of the core positions due to the finite cell size.

Making inverse FFTs of all of those peaks of each mode with α = 0.0001, we obtained the spatial correlation between the five cores' motion for each mode, as shown in Fig. 3. The trajectories of the orbiting cores in one cycle period (2π/ω) of gyration starting from about 100 ns are compared for all of the modes, along with the corresponding profiles of the *Y* components. The *Y* component profiles contrast markedly between the modes. The shapes of all of the trajectories are eccentric along either the *x* or *y* axis, because the strong exchange interaction between the antivortices and the neighboring vortices is additionally employed in such a connected-disk model, with its high asymmetry between the *x* and *y* axes. The most noteworthy feature is the fact that the collective core motions show a standing-wave form with a different overall wavelength for a given mode. The orbiting radii of the five cores for a given mode are symmetric with respect to the center of the whole system (i.e., at V$_3$) and are also completely pinned at the imaginary points at both ends, denoted AV$_0$ and AV$_6$, as reported in Refs.[20, 24]. The lower $\omega_1$, $\omega_2$, and $\omega_3$ modes and the higher $\omega_4$ and $\omega_5$ modes are quite different in terms of their standing-wave forms. For the $\omega_1$ mode, all of the cores move in phase, while for the $\omega_2$ mode, the core of $V_3$ acts as a node, and for the $\omega_3$ mode, the two antivortex cores act as nodes. For the $\omega_4$ and $\omega_5$ modes, the two antivotices are highly excited relative to the three vortices. Also, the antivortices' core motions are permeated into the bonding axis, thus resulting in higher energy states due to their strong exchange interaction with the neighboring vortices. The difference between the individual modes can be understood by the relative phases between the vortex's and antivortex's effective magnetizations, <**M**>, induced by their own core shifts and consequently by their dynamic dipolar interaction, as discussed in Ref. 20. The relative phases of the <**M**> between the neighboring core gyrations determine the mode's average dynamic dipolar energy. However, the strong exchange interaction in such a connected thin-film strip must also be taken into account. Due not only to the unknown potential wells of the antivortices but also the complex asymmetric exchange interaction terms between the vortices and antivortices, it is difficult, in such V-AV lattices, to separately extract the individual contributions of the exchange and dipolar interaction terms to the individual coupled core motions in each mode. A distinct feature of the V-AV coupled lattices is that the strong exchange coupling between the dynamic vortex and antivortex motions dominantly contributes to each mode and differently contributes to the individual modes, as evidenced by the presence of the higher frequency modes that would not be found in vortex-state arrays.[20]

Next, on the basis of the above findings, we extended our study to longer 1D chains, as shown in Fig. 4(a). Here we consider two different polarization orderings with the same CCW chirality for all of the individual vortices: antiparallel and parallel orderings represented by ($p_V$, $p_{AV}$) = (+1, -1) and ($p_V$, $p_{AV}$) = (+1, +1), respectively. For such V-AV finite lattices, the thickness was set to 40 nm to make them more stable than in a thinner strip and we applied a static local field of 200 Oe in the -y direction only in the left-end disk. The other simulation conditions were the same as those for the earlier model shown in Fig. 1. From the FFTs of the *x* components of all of the 25 core positions, we obtained dispersion curves in the reduced zone scheme for the antiparallel and parallel polarization orderings, as shown in Fig. 4(b). Here, we assume that the interdistance between the neighboring vortices and antivortices is equal to the average value, $\bar{d}_{int}/2 = 121$ nm, where $\bar{d}_{int}$ is the lattice constant of a unit basis for V-AV lattices, though there are slight differences in the cores' distances in the center and end regions. The dispersions were asymmetric with respect to the wavenumber *k*=0, because the initial core motion was excited only in the left vortex and was then propagated toward the +*x* direction. Therefore, the modes with positive group velocities were relatively strong as compared with those with the negative group velocities. In such band structures, there are two distinct higher- and lower-frequency branches, as expected from the two different types of standing-wave forms found in the earlier V-AV model shown in Fig. 1. It is noteworthy that such dispersions are quite analogous to collective gyrations in the *acoustic* and *optical branch*, respectively, as observed in only vortex-state lattices consisting of alternating different materials [24] or in diatomic lattice vibrations. This indicates that the present 1D V-AV lattice, even with a single soft magnetic material, acts as a bi-material medium. We also note that each branch consists of quantized flat-shaped local modes. In the FFTs of the coupled V-AV gyrations, we applied a periodic boundary condition: as a bi-object array, the wave number is set to $k=\pi m/N\bar{d}_{int}$, where *N* is the number of the unit basis, and *m* is an arbitrary integer in the $-\pi/\bar{d}_{int} \leq k < \pi/\bar{d}_{int}$ constraint.[40] At each of the *N*-discrete *k* values, there are two corresponding frequencies, thus leading to 2*N* normal modes.[24,40,41]

In both polarization orderings, the bandgap between the two branches is almost the same, ~0.25 GHz, and the lowest frequency of the higher branches is ~1.23 GHz. The band width of the lower branch is as wide as ~0.9 GHz, due to the strong exchange-dipole interaction between vortices and antivortices in such connected-disk arrays. The overall shape of the lower branch is concave up; that is, the frequency is lowest at $k=0$ and highest at the first Brillion zone boundary, $k=k_{BZ}=\pi/\bar{d}_{int}$, for both polarization orderings. However, the shape of the higher branch varies according to the polarization ordering: concave down for the antiparallel polarization ordering, and almost flat for the parallel one. In the lower branch, as $k$ approaches the $k_{BZ}$, the $\omega$ value reaches the angular eigenfrequency $\omega_0$ of isolated Py disks of the given dimensions (here $\omega/2\pi = 0.97$ GHz). This result reveals that all of the antivortices act as nodes in the standing-wave form, and thus do not contribute to the lower band at $k = k_{BZ}$ but dominantly contribute to the higher band at $k = k_{BZ}$. The flat higher band for the parallel polarization ordering is owed to the fact that the dynamic interaction energy averaged over one cycle of gyration is almost equal for the entire $k$ range. The $\omega$ value in the higher flat band would provide the angular eigenfrequency of a virtual system composed of isolated antivortices (here ~1.23 GHz).

In fact, this connected-disk system is analogous to a nanostrip with round-shaped width modulations. Thus, from the technological perspective, as with domain-wall motions in a given nanostrip, gyration-signal propagation through alternating V-AV lattices in a planar-patterned nanostrip can be used as an information carrier. In order to estimate the gyration-signal propagation speeds, we plotted the displacements of the individual cores from their own center positions in the whole system with time, as seen in Fig. 4(c). This figure clearly indicates that the excited gyrations propagate well through the entire connected-disk nanostrips. The two slopes noted by the white lines compare the substantial propagation speed difference between the parallel and antiparallel polarization orderings. The speed for the parallel (antiparallel) polarization ordering is 0.83 (0.64) km/s, representing a 30% increase relative to the antiparallel ordering. It is known that in the case of dipolar-coupled arrays composed only of vortex-state disks, the gyration propagation for the antiparallel polarization ordering is faster than that for the parallel polarization ordering, owing to the stronger dynamic dipolar interaction, as determined by the relative rotation senses of the two core gyrations.[18,20] However, for the cases of V-AV lattices the faster propagation speed for the parallel ordering than for the antiparallel ordering cannot be explained only in terms of the dynamic dipolar interactions between the vortices and antivortices. Although we could not understand why the exchange interaction affects the faster gyration propagation for the parallel polarization ordering, we could confidently conclude that the exchange interaction between the vortices and antivortices gives rise to a faster propagation speed for either polarization ordering in the V-AV lattices than in physically separated disk arrays of only vortex states. We also note that the gyration-signal propagation through the antivorticies is comparable to or faster than the domain-wall speeds in nanostrips.[42] This is another antivortex-mediated mechanism for faster gyration-signal propagation in continuous nanostrips.

Next, in order to control the observed two-branch band structure and the gyration-signal propagation speed in such V-AV lattices, we applied perpendicular bias fields $H_z$ of different strength and direction. It is known that the eigenfrequency of a gyrotropic mode varies with $H_z$, as expressed by $\omega = \omega_0(1 + pH_z/H_s)$, with $\omega_0$ the angular eigenfrequency at $H_z = 0$, and $H_s$ the perpendicular field for the saturation of a given system's magnetization.[43,44] In our numerical calculations, the eigenfrequency of an isolated antivortex was estimated to be linearly proportional to $H_z$ as well (not shown here). Thus, it is interesting to examine how coupled gyrations in such V-AV lattices, and consequently how the resultant band structure, varies with $H_z$. In our further simulations, we excited vortex and antivortex gyrations by the same method as described earlier, but under the application of different values of $H_z$ at intervals of 1 kOe in the $H_z = -3 - +3$ kOe range. Figure 5 compares the contrasting band structures for the indicated $H_z$ values. The band widths of the lower and higher bands and their bandgaps markedly vary with $H_z$ and differ from the parallel to antiparallel polarization ordering. For example, at $H_z = -3.0$ kOe, in the case of $(p_V, p_{AV}) = (+1, +1)$, the higher flat band becomes stronger and more flat over a wide range of $k$, and the lower band becomes weak, while in the case of $(p_V, p_{AV}) = (+1, -1)$, the higher flat band becomes very weak and the lower band becomes relatively strong. Also in the case of $(p_V, p_{AV}) = (+1, +1)$, the lower band width increases with $H_z$, while in the case of $(p_V, p_{AV}) = (+1, -1)$ it increases with $H_z$ until $H_z = 1$ kOe and then decreases again. From the band structures obtained with the simulations, we plotted the angular frequencies at $k=k_{BZ}$ (denoted $\omega_{BZ}$ hereafter) for the lower and higher bands. As for $(p_V, p_{AV}) = (+1, +1)$, the highest energy of the lower band and the lowest energy of the higher band are linearly proportional to $H_z$ in the given $H_z$ range, with slopes of 0.12 GHz/kOe and 0.21 GHz/kOe, respectively. In fact, those values of $\omega_{BZ}$ for the higher and lower bands correspond to the eigenfrequencies of the isolated antivortex and vortex, respectively. Thus, the linear dependences of $\omega_{BZ}$ on $H_z$ for both bands are associated with the variation of the eigenfrequencies of isolated vortices and antivotices with $H_z$, as expressed by $\omega = \omega_0(1+ pH_z / H_s)$. Using this equation, linear fits to the data yield the two fitting parameters: $\omega_0 = 0.95 \pm 0.002$ (GHz), $H_s = 8.2 \pm 0.08$ (kOe) for the lower band and $\omega_0 = 1.2 \pm 0.004$ (GHz), $H_s = 5.9 \pm 0.01$ (kOe) for the higher band. These two values of $\omega_0 = 0.95$ and 1.2 GHz are close to the eignfrequecis of isolated vortices and antivortices. The larger slope for the higher band is the result of the higher $\omega_0$ and lower $H_s$.

In the case of $(p_V, p_{AV}) = (+1, -1)$, on the other hand, the $\omega_{BZ}$ of the higher band is linearly decreased and then, after crossing about $H_z = 1$ kOe, increased. The $\omega_{BZ}$ of the lower band, meanwhile, shows exactly the reverse effect. The higher (lower) band in the $H_z < 1$ kOe range seems to follow the relation $\omega = \omega_0(1 - H_z / H_s)$ ($\omega = \omega_0(1 + H_z / H_s)$), whereas in the $H_z > 1$ kOe range, $\omega = \omega_0(1 + H_z / H_s)$ ($\omega = \omega_0(1 - H_z / H_s)$). These results reveal that in the $H_z < 1$ kOe ($H_z > 1$ kOe) range, antivortices play a dominant role in the higher (lower) band and that vortices do the same in the lower (higher) band. That is why the band-structure variation with $H_z$ is rather more complicated in the case of $(p_V, p_{AV})=(+1, -1)$ than in the case of $(p_V, p_{AV}) = (+1, +1)$. Applications of $H_z$ to such V-AV coupled lattices allow for manipulation of the relative roles of the vortex and antivortex in their coupled motions. Owing to the combined strong exchange and dipolar coupling between the neighboring vortices and antivortices, the vortex and antivortex's roles would not be discriminated separately in those band structures within most $k$ ranges.

In addition, we also estimated the gyration-signal propagation speeds versus $H_z$ for both polarization ordering cases. Figure 6(b) shows that the resultant propagation speed is linearly proportional to $H_z$ in the given $H_z$ range for the parallel polarization ordering (open squares), but that for the antiparallel ordering (open circles), the speed decreases with increasing $|H_z|$ and, thereby, is at the maximum at $H_z=0$. The speed difference ratio between the parallel and antiparallel orderings increases markedly with $H_z$, as shown in Fig. 6(c). For example, the difference ratio increases to ~135% at $H_z = 3$ kOe, compared with 28% at $H_z=0$ and 2% at $H_z =-3$ kOe. This remarkable variation of the speed difference ratio with $H_z$ is the result of increases in the intrinsic eigenfrequencies of both the vortex and antivortex with $H_z$ for ($p_V$, $p_{AV}$) = (+1, +1) and the result of decreases in the antivortex eigenfrequency for ($p_V$, $p_{AV}$) = (+1, -1). These results are very promising from the technological point of view, owing to the following several advantages of this signal-propagation mechanism and its applicability to any potential spin-based signal-processing devices: 1) such V-AV lattices can be made of simple, single-material, round-shaped modulated patterned thin-film nanostrips; 2) the parallel polarization ordering between neighboring vortices and antivortices is readily available by application of a strong magnetic field perpendicularly to the strip plane; 3) such ultrafast gyration-propagation speed is caused by the combined strong exchange-dipolar coupling between the neighboring vortices and antivortex, and accordingly the propagation speed is higher than 1 km/sec and controllable by application of $H_z$.

In summary, we studied the dynamics of coupling between vortices and antivortices in their alternating 1D periodic arrays. Standing-wave discrete modes and their dispersion relations were found to have characteristics of two branches: lower and higher bands. These lower and higher bands are strongly influenced by the polarization ordering between the neighboring vortex and antivortex, and thereby are controllable. We found faster gyration-signal propagations in continuous nanostripes composed of alternating vortex and antivortex lattices for the parallel polarization ordering than for the antiparallel one; both are much faster than only-vortex-state arrays. Also, the band structures and gyration-propagation speed are markedly variable by means of the perpendicular bias field. This work provides an additional mechanism of ultrafast gyration-signal propagation via exchange-dipole interaction, and offers, moreover, a fundamental understanding of vortex and antivortex coupled dynamics in one-dimensional magnonic crystals.

This research was supported by the Basic Science Research Program through the National Research Foundation of Korea funded by the Ministry of Science, ICT & Future Planning (Grant No. 2014001928).

# References

†These two authors equally contributed to the work.

*Author to whom all correspondence should be addressed; electronic mail: sangkoog@snu.ac.kr

**FIGURE 1.**

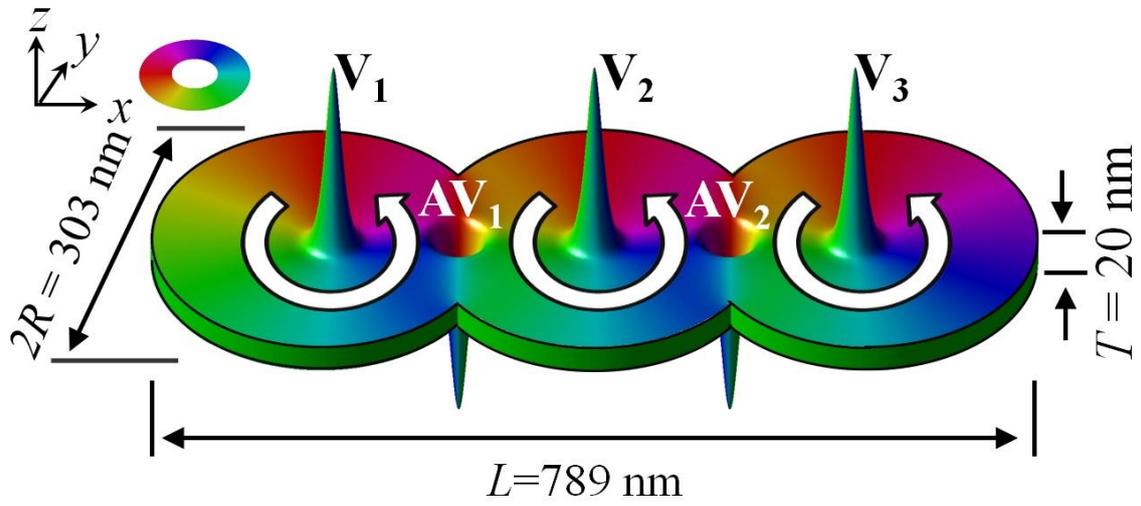

FIG. 1. (color online) Model geometry of connected triple-disk structure. The individual disks have an equal diameter $2R = 303$ nm, a thickness $T = 20$ nm, and a disk center-to-center distance $D_{int} = 243$ nm. The color and height display the in-plane magnetization and out-of-plane magnetization components, respectively. The chirality of the three vortices is CCW, as indicated by the white arrows, and the polarizations of the three vortices and two antivortices are upward and downward, respectively.

**FIGURE 2.**

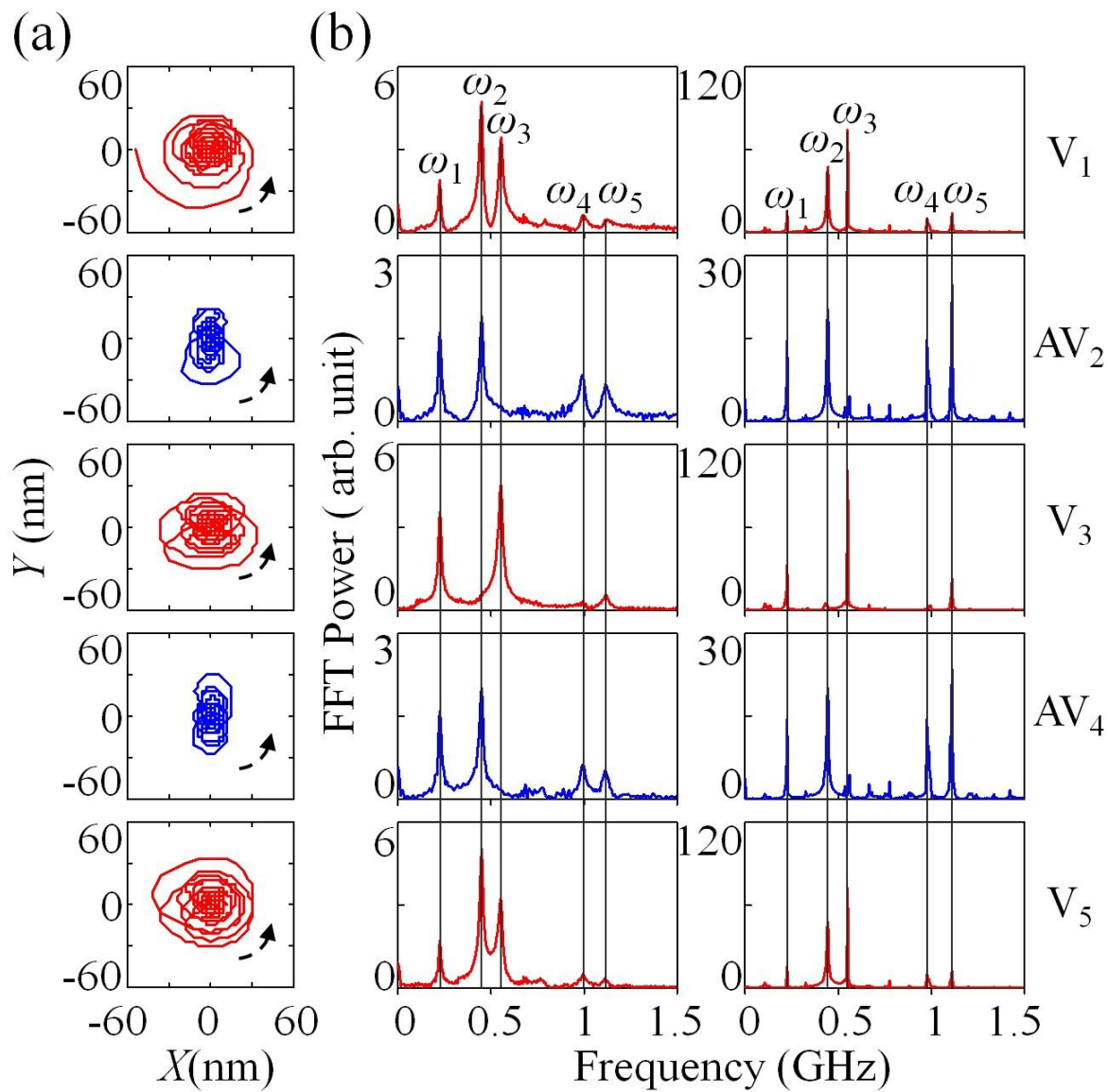

FIG. 2. (color online) (a) Trajectories of gyration motions of individual vortex cores (red lines) and antivortex cores (blue lines). (b) FFTs of $x$ components of individual core-position vectors from their own center positions. The left (right) column is the result with $\alpha = 0.01$ (0.0001). The five major peaks are denoted $\omega_i$ where $i = 1, 2, 3, 4$, and 5, and are marked by the gray vertical lines.

**FIGURE 3.**

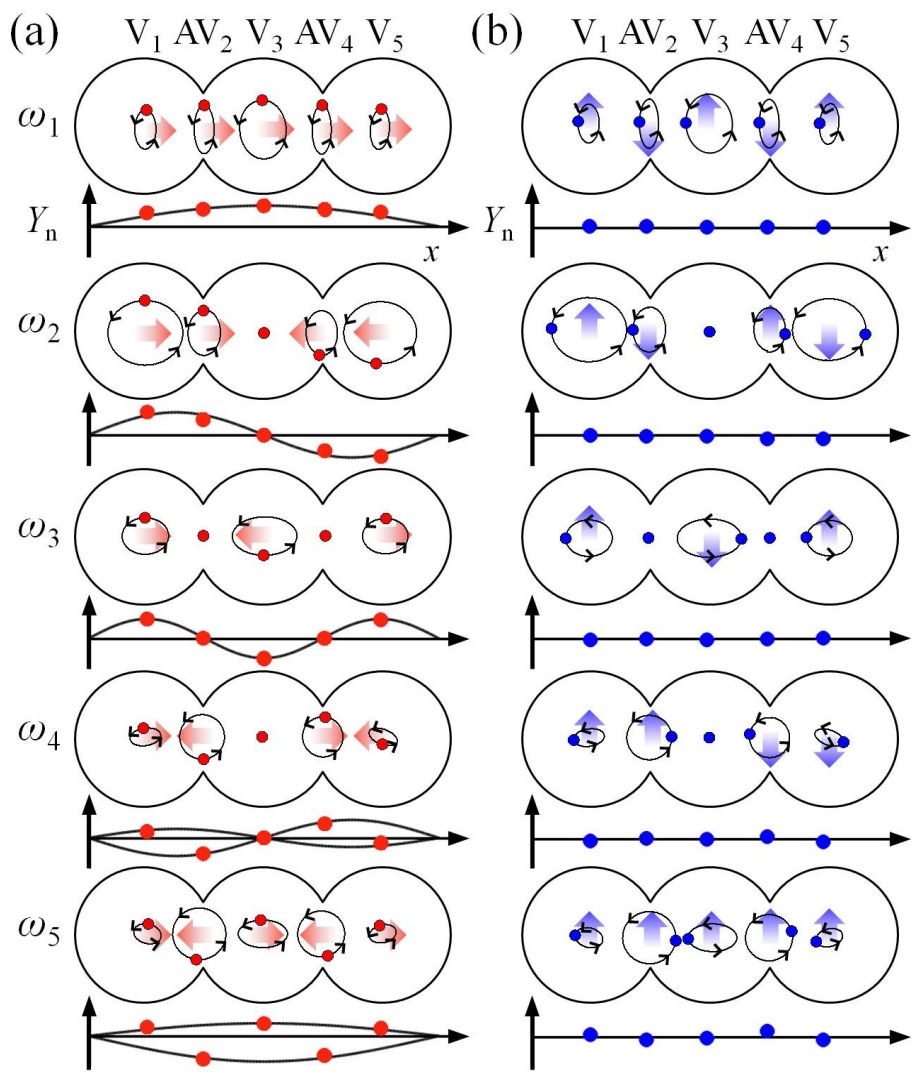

FIG. 3. (color online) Spatial distributions of individual core positions for five different collective motions (modes). The left and right columns correspond to the core positions taken at $T$ and $T/4$ after about 100 ns, which are aligned along the $y$ and $x$ axes, respectively, as marked by the red and blue dots on the trajectories. The core motions' trajectories are magnified for clear comparison (though the magnifications are different for the different modes). The wide arrows represent the directions of the effective magnetizations <**M**> induced by their own core shifts.

**FIGURE 4.**

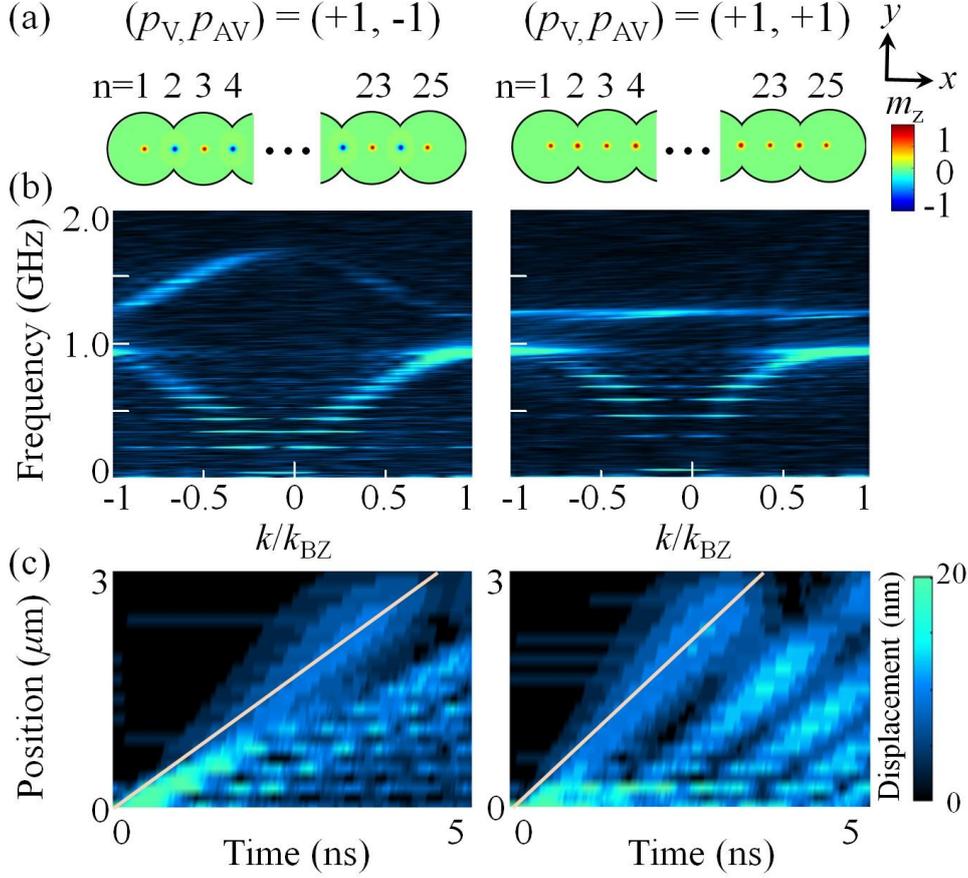

FIG. 4. (color online) (a) Model geometry of 1D chains comprising 13 vortices and 12 antivortices between the neighboring vortices. Odd (even) index numbers from 1 to 25 represent vortices (antivortices). The in-plane curling magnetization of all of the vortices is CCW. The red (blue) dots display upward (downward) core orientation. (b) Dispersions of collective V-AV gyration modes in given V-AV chains. FFT spectra were obtained from the $x$ components of the core positions using the zero-padding technique for a sufficient $k$-space resolution of $1.4 \times 10^6$ m$^{-1}$. The results are plotted in the reduced zone scheme. (c) Plane-view representations of core displacements with respect to time and distance in entire chain. The zero position is defined as the first vortex-core position in the initial state. The color bar indicates the displacement of each vortex core or antivortex core from each center position.

**FIGURE 5.**

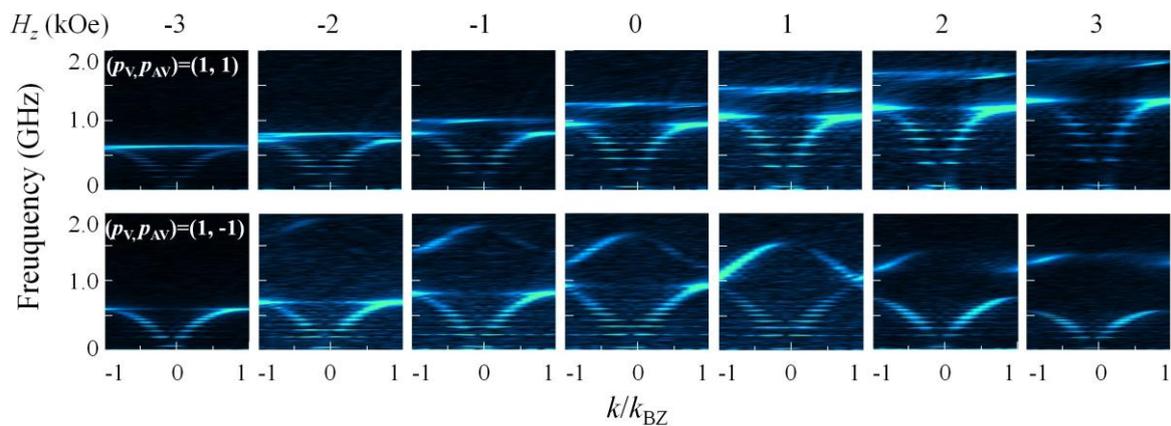

FIG. 5 (color online) Two-branch band-structure variation with $H_z$ for both antiparallel (($p_V$, $p_{AV}$) = (+1, -1)) and parallel (($p_V$, $p_{AV}$) = (+1, +1)) polarization orderings in V-AV lattice shown in Fig. 4(a)

**FIGURE 6.**

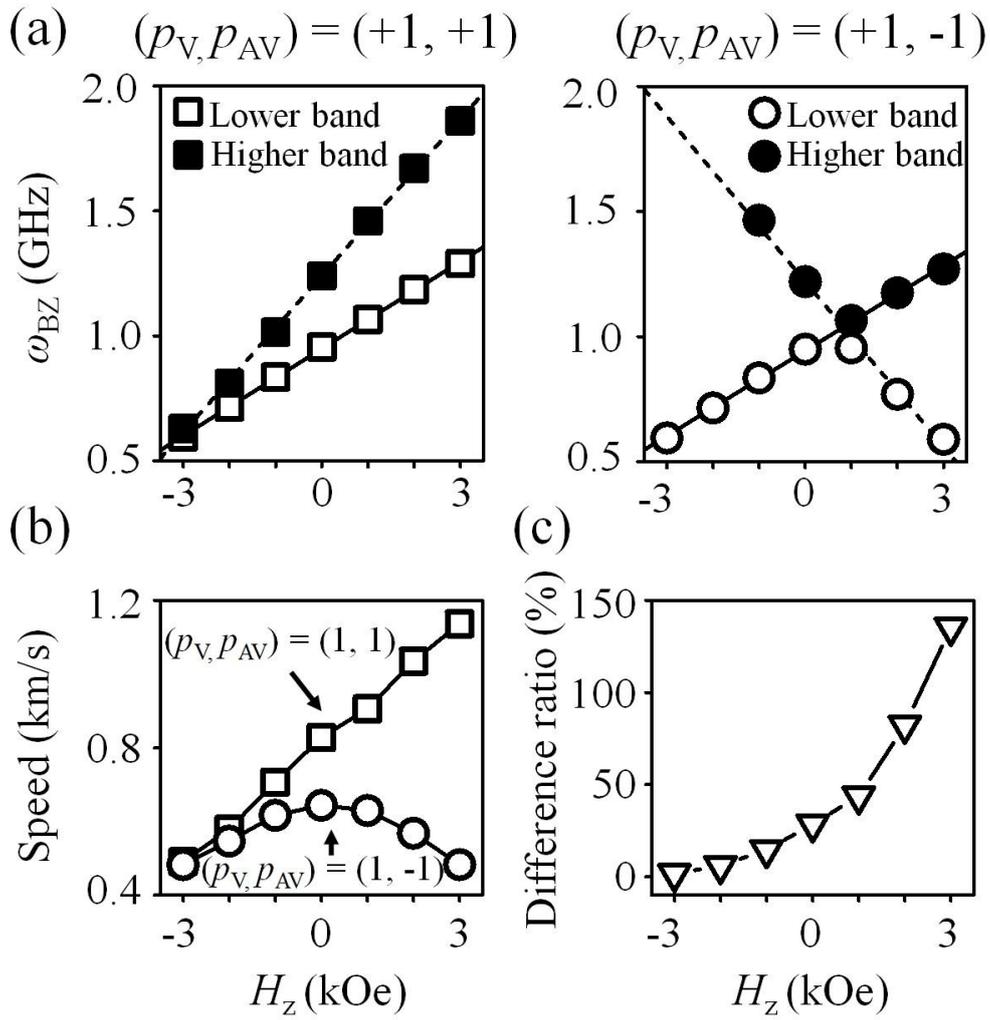

FIG. 6 (a) Plots of angular frequency at $k=k_{BZ}$ for lower and higher bands in both polarization ordering cases. The solid and dashed lines represent the results of linear fits to the data. (b) Gyration-signal propagation speed in both polarization ordering cases and (c) their difference ratio.

23